\documentclass[a4paper]{aa}
\usepackage{psfig}
\usepackage{epsfig}
\usepackage{graphicx}

\def\ltsima{$\; \buildrel < \over \sim \;$}
\def\lsim{\lower.5ex\hbox{\ltsima}}
\def\gtsima{$\; \buildrel > \over \sim \;$}
\def\gsim{\lower.5ex\hbox{\gtsima}}

\def\sgr {SGR~1627--41}
\def\igr {IGR~J16358--4726}

\begin{document}


\title{XMM-Newton observations of the Soft Gamma Ray Repeater
 \sgr\  in a low luminosity state}

\author{S. Mereghetti\inst{1}
        \and P. Esposito\inst{1,2}
        \and A. Tiengo\inst{1,2}
        \and R.~Turolla\inst{3}
        \and S.~Zane\inst{4}
        \and L.~Stella\inst{5}
        \and G.L.~Israel\inst{5}
        \and M.~Feroci\inst{6}
        \and A.~Treves\inst{7}
}

 \institute { {1) INAF  - Istituto di Astrofisica Spaziale e Fisica
Cosmica Milano, via Bassini 15, I-20133 Milano, Italy}
\\
{2) Universit\`a degli Studi di Milano, Dipartimento di Fisica,
via Celoria 16, I-20133 Milano, Italy}
\\
{3) Universit\`a di Padova, Dipartimento di Fisica, via Marzolo 8,
I-35131 Padova, Italy}
 \\
{4) Mullard Space Science Laboratory, University College London,
Holmbury St. Mary, Dorking Surrey, RH5 6NT, UK}
\\
 {5) INAF - Osservatorio Astronomico di Roma, via
Frascati 33, I-00040 Monteporzio Catone, Roma, Italy}
\\
 {6)  INAF  - Istituto di Astrofisica Spaziale e Fisica
Cosmica Roma, via Fosso del Cavaliere  100, I-00133 Roma, Italy}
\\
 {7) Universit\`a degli Studi dell'Insubria,
 via Valleggio 11, I-22100 Como, Italy}
 }

\offprints{S. Mereghetti (sandro@iasf-milano.inaf.it)}

 \date{Received  15 September 2005 ; Accepted: 11 January 2006 }

\authorrunning{S. Mereghetti et al.}

\titlerunning{{ \sgr }}

\abstract{The sky region containing the soft gamma-ray repeater
\sgr\ has been observed three times with XMM-Newton in February
and September 2004. \sgr\ has been detected with an absorbed flux
of $\sim$9$\times$10$^{-14}$ erg cm$^{-2}$ s$^{-1}$ (2-10 keV).
For a distance of 11 kpc, this corresponds to a luminosity of
$\sim$3$\times$10$^{33}$ erg s$^{-1}$, the smallest ever observed
for a Soft Gamma Repeater and possibly related to the long period
of inactivity of this source. The observed flux is smaller than
that seen with Chandra in 2001-2003, suggesting that the source
was still fading and had not yet reached a steady quiescent level.
The spectrum is equally well fit by a steep power law (photon
index $\sim$3.2) or by a blackbody with temperature kT$\sim$0.8
keV. We also report on the INTEGRAL transient \igr\ that lies at
$\sim$10$'$ from \sgr . It was detected only in September 2004
with a luminosity of $\sim4\times10^{33}$ erg s$^{-1}$ (for d=7
kpc), while in February 2004 it was at least a factor 10 fainter.
  }

\maketitle

\section{Introduction}
\label{sect:intro}

\sgr\ is one of the four confirmed Soft Gamma-ray Repeaters (SGRs)
that are currently known. According to the widely  accepted
magnetar model (Duncan \& Thompson 1992, Thompson \& Duncan 1995),
these sources are isolated neutron stars in which the high-energy
emission is powered by ultra-strong magnetic fields
(B$\sim$10$^{14}$-10$^{15}$~G). The distinctive characteristic of
SGRs is the emission, during sporadic periods of activity, of
short bursts ($<$ 1 s) of hard X--rays with super-Eddington peak
luminosity  L$\sim$10$^{40}$--10$^{41}$ erg s$^{-1}$. Persistent
(i.e. non-bursting) emission is also observed from SGRs in  the
soft X--ray range ($<$10 keV), with typical luminosity of
$\sim$10$^{35}$ erg s$^{-1}$. Periodic pulsations at several
seconds, reflecting the neutron star rotation, are observed in
three SGRs. Occasionally, SGRs also emit energetic (giant) flares
with luminosity from $\sim$10$^{43}$ erg s$^{-1}$ up to 10$^{47}$
erg s$^{-1}$.  For a review of the properties of these sources see
Woods \& Thompson (2004).

It is interesting to study the relation between the properties of
the persistent X--ray emission and the level of SGR bursting and
flaring activity. In fact the persistent soft X--rays are thought
to consist, at least in part, of thermal emission from the neutron
star surface, whose properties, e.g. temperature and
magnetization, can be influenced by the largely non-thermal
phenomena responsible for the bursts.

The 1-10 keV luminosity of the two SGRs which have been more
active in recent years, SGR 1806--20 and SGR 1900+14, displayed
only moderate (factor $\sim$2) long term variations around average
values of $\sim5\times10^{35}$d$^2_{15}$ erg s$^{-1}$ and
$\sim1\times10^{35}$d$^2_{10}$ erg s$^{-1}$
respectively\footnote{we indicate with d$_N$ the distance in units
of N kpc} (e.g., Mereghetti et al. 2005, Woods et al. 2001). The
SGR in the Large Magellanic Cloud, SGR 0526--66, has a similar
luminosity of $\sim10^{36}$d$^2_{55}$ erg s$^{-1}$ (Kulkarni et
al. 2003), despite no bursts have been detected from this source
since 1984\footnote{some bursts might have been missed in the time
interval 1985--1991 due to the lack of suitable detectors in
operation (see Woods \& Thompson 2004)}. The only difference with
respect to the two more active SGRs mentioned above is that its
spectrum is much softer, requiring a power law photon index larger
than $\sim$3 (Kulkarni et al. 2003).

\sgr\,  was discovered  in 1998, when more than 100 bursts were
observed with different satellites (CGRO, Woods et al. 1999;
Ulysses, Hurley et al. 1999; Wind, Mazets et al. 1999;  RXTE,
Smith et al. 1999; BeppoSAX, Feroci et al. 1998). No other bursts
from this source have been reported to date. Its soft X-ray
counterpart was identified with BeppoSAX in 1998 at a luminosity
level of $\sim$10$^{35}$d$^2_{11}$  erg s$^{-1}$ (Woods et al.
1999). Subsequent observations, carried out over a time span of
five years with BeppoSAX, ASCA and Chandra, showed a monotonic
decrease in the luminosity, interpreted  as evidence for cooling
of the neutron star surface after the deep crustal heating that
occurred during the 1998 period of SGR activity (Kouveliotou et
al. 2003). The latest Chandra observation (March 2003) yielded an
X-ray  flux consistent with that measured in September 1999,
suggesting that the luminosity of \sgr\ settled at its
''quiescent'' level of $\sim$4$\times$10$^{33}$d$^2_{11}$ erg
s$^{-1}$. To further study the luminosity evolution of \sgr\ we
observed it with XMM-Newton in September 2004. The SGR was also
serendipitously detected in two other XMM-Newton observations
pointed on the transient IGR J16358--4726 (Patel et al. 2004),
which lies at an angular distance of $\sim$10$'$. We report here
also the results of these serendipitous detections, as well as a
reanalysis of the BeppoSAX observations. For completeness, we
report also the results of the three XMM-Newton observations for
IGR J16358--4726.

\section{Data Analysis and Results}

\subsection{SGR 1627--41: on-axis data}

Our observation of \sgr\ was carried out on 2004 September 22 and
lasted about 52 ks. Here we report on the results obtained with
the three EPIC CCD cameras (Turner et al. 2001, Str\"{u}der et al.
2001) since the target was too faint for the RGS instruments. All
the CCDs were operated in Small Window mode, yielding time
resolution of 6 ms (PN) and 0.3 s (MOS), and the medium thickness
filter was used.

A faint, but statistically significant, source was detected at the
coordinates of \sgr\ (Wachter et al. 2004) in all the EPIC
cameras. Its background-subtracted count rates were 2.5$\pm$0.4
and 3.0$\pm$0.4 counts ks$^{-1}$ in the two MOS cameras (with net
exposure times of 50.3 ks each) and 9.4$\pm$1.2 counts  ks$^{-1}$
in the PN (net exposure 36.2 ks). We extracted the source spectra
using a circular extraction region with  25$''$ radius and the
background from source free regions of the same observation. The
spectra were rebinned to have at least 30 counts per channel. The
spectra from the three cameras were fitted together, with the
appropriate response matrices, to simple models. Although the
small number of counts allowed us to carry out only a limited
spectral analysis, there is clear evidence for a rather soft
spectrum. Both a steep power law (photon index $\Gamma=3.7\pm0.5$)
and a blackbody with temperature kT$_{BB}=0.8^{+0.2}_{-0.1}$ keV
gave acceptable fits. The best fit power-law parameters are given
in Table 1, where for comparison with previous work, we also
report the values obtained by keeping the absorption fixed at the
average value of the other observations, N$_H$=9$\times10^{22}$
cm$^{-2}$ (Kouveliotou et al. 2003). The blackbody fit gives an
absorption N$_H$=(6.8$^{+2.7}_{-2.3}$)$\times10^{22}$ cm$^{-2}$,
an observed 2-10 keV flux of (7.5$\pm$1.5)$\times10^{-14}$ erg
cm$^{-2}$ s$^{-1}$, and, for an assumed distance of 11 kpc, an
emitting radius of 0.2$\pm$0.1 km.

We searched for periodic pulsations, using a small extraction
region in order to reduce the fraction of background counts (45\%
of the total counts within the adopted 10$''$ radius circle). The
pulsations were not detected, but due to the limited statistics
the upper limits on the source pulsed fraction are not
particularly constraining. No evidence for bursts or other flux
variations was seen.

\subsection{SGR 1627--41: off-axis data}

Two XMM-Newton observations of IGR J16358--4726 were carried out
on 2004 February 16 and September 4, with a duration of about 30
ks each.  In both observations \sgr\ was detected with EPIC at an
off-axis angle of 9.6$'$, where, owing to the vignetting effect,
the effective area is about half of the on-axis one. The
background-subtracted PN count rates were 5.4$\pm$0.8 and
4.2$\pm$0.6  counts ks$^{-1}$, respectively.  The counts
statistics was too poor for a spectral analysis, but the source
hardness ratios were consistent with the ones of the on-axis
observation. We therefore computed the source fluxes given in
Table 1 by fixing N$_H=9\times10^{22}$ cm$^{-2}$ and assuming the
same photon index as the on-axis observation.

\begin{figure}[th!]
\begin{center}
\psfig{figure=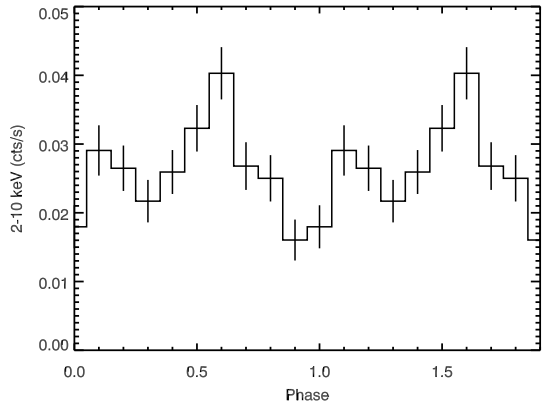,width=6cm,angle=90}
 \caption{Light curve of IGR J16358--4726 folded at 5880 s. The light curve
 is the sum of PN and MOS data and the background has been subtracted.
 The $\chi^2$ statistics corresponds to a probability of 2$\times10^{-5}$ that
 the source is constant.}
\end{center}
\end{figure}

\begin{table*}[htbp]
\begin{center}
  \caption{Spectral results on \sgr .  The
errors of the BeppoSAX and XMM-Newton values are at 1$\sigma$ for
a single parameter of interest, and, for the fluxes, take into
account both the statistical and spectral uncertainties. The ASCA
and Chandra values are from Kouveliotou et al. (2003). }

    \begin{tabular}[c]{lccccc}
\hline
Satellite & Observation  Date   & N$_H$           &  photon &  Observed flux$^{(a)}$ & Unabsorbed flux$^{(a)}$ \\
            &        & 10$^{22}$ cm$^{-2}$ & index    & 10$^{-12}$ erg cm$^{-2}$ s$^{-1}$    &  10$^{-12}$ erg cm$^{-2}$ s$^{-1}$   \\
\hline
BeppoSAX & 1998 August 6-7      & $10\pm1 $ & $2.6 \pm 0.3$ & $2.5 \pm 0.2$ & $5.9^{+1.4}_{-1.0} $ \\
BeppoSAX &1998 September 16 & $10\pm2$ & $2.8 \pm 0.4$ & $1.9 \pm 0.2$ & $4.6^{+1.3}_{-0.9}$ \\
BeppoSAX &1999 August 8-10     & $5 \pm 2$ & $1.9^{+0.4}_{-0.7}$ & $0.7 \pm 0.1$ & $1.1 \pm 0.3$ \\
BeppoSAX & 2000 September 5-7  & $8^{+5}_{-4}$ & $3.0^{+1.3}_{-0.9}$ & $0.4^{+0.2}_{-0.1}$ & $1.1\pm0.4 $ \\
XMM-Newton & 2004 September 22 & $11.4^{+4.3}_{-1.9}$ & $3.7 \pm 0.5$ & $0.09 \pm 0.02$ & $0.32 \pm 0.09$ \\
\hline
BeppoSAX & 1998 August 6-7    & 9$^{(b)}$ & $2.4 \pm 0.1$ & $2.6 \pm 0.1$ & $5.2 \pm 0.3$ \\
BeppoSAX &1998 September 16   & 9$^{(b)}$ & $2.6 \pm 0.2$ & $1.9 \pm 0.1$ & $4.2 \pm 0.3$ \\
ASCA    & 1999 February 26-28 & 9$^{(b)}$ & $3.24 \pm 0.24$ & 1.07 $\pm$ 0.16 & $2.76 \pm 0.41$    \\
BeppoSAX &1999 August 8-10    & 9$^{(b)}$ & $2.5 \pm 0.3$ & $0.7\pm 0.1$ & $1.5 \pm 0.2$ \\
BeppoSAX & 2000 September 5-7 & 9$^{(b)}$ & $3.3 \pm 0.5$ & $0.4 \pm 0.1$ & $1.1 \pm 0.2$\\
Chandra & 2001 September 30  &  9$^{(b)}$ & $2.17 \pm 0.30$ & $0.14 \pm 0.02$ & $0.267 \pm 0.04$  \\
Chandra & 2002 August 19     &  9$^{(b)}$ & $2.95 \pm 0.36$ & $0.11 \pm 0.02$ &  $0.266 \pm 0.04$ \\
XMM-Newton & 2004 February 16 & 9$^{(b)}$ & 3.2$^{(b)}$   & $0.12 \pm 0.02$ & $0.32 \pm 0.06$  \\
XMM-Newton & 2004 September 4 & 9$^{(b)}$ & 3.2$^{(b)}$   & $0.10 \pm 0.02$ & $0.26 \pm 0.04$ \\
XMM-Newton & 2004 September 22 & 9$^{(b)}$& $3.2 \pm 0.3$   & $0.09 \pm 0.01$ & $0.23 \pm 0.04$ \\
\hline

\end{tabular}
\end{center}
\begin{small}
$^{(a)}$ in the 2-10 keV energy range \\
$^{(b)}$ fixed value  \\
\end{small}
\label{obslog}
\end{table*}

\subsection{IGR J16358--4726}

The transient \igr\ was  discovered in the hard X-ray band ($>$ 15
keV) with INTEGRAL in March 2003 (Revnivtsev et al. 2003). Lying
at  a small angular distance from \sgr, this source was
serendipitously present in several observations of the soft
repeater. During the March-April 2003 outburst, periodic
pulsations at 5880 s were discovered with Chandra (Patel et al.
2004). It is not clear yet whether  they correspond to the spin
period of a neutron star in a High Mass X-ray Binary or the
orbital period of a Low Mass X-ray Binary.

\igr\ was detected only in the two observations of September 2004.
The best data  were obtained on September 4, when the source was
detected on-axis, with background-subtracted count rates of
4.4$\pm$0.4, 4.3$\pm$0.4 and 18.9$\pm$1.0 counts ks$^{-1}$ in the
MOS1, MOS2 and PN, respectively. The spectral analysis of these
data, carried out as described in Section 2.1, yielded the
following best fit parameters for an absorbed power law model:
$\Gamma$=1.5$\pm$0.5, N$_H$=(20$\pm$5)$\times10^{22}$ cm$^{-2}$,
observed 2-10 keV flux = (3.1$\pm$0.6)$\times10^{-13}$ erg
cm$^{-2}$ s$^{-1}$. The background subtracted light curve folded
at 5880 s (Fig.~1), shows that the periodic modulation, observed
with Chandra when the source was a factor ~200 more luminous, is
present also in this lower intensity state.

In the subsequent observation (September 22)  \igr\ was detected
only in the two MOS cameras, since its off-axis position was
outside the region covered by the PN in Small Window mode.
Although the limited statistics ($\sim$200 net counts in total) do
not allow us to see the pulsations or to significantly constrain
the spectral parameters, the MOS count rates were consistent with
the intensity and spectral shape measured 18 days earlier.

Adopting a power law spectrum with photon index $\Gamma$=1.3 and
N$_H$=2$\times10^{23}$ cm$^{-2}$, in order to compare with the
previous detections with ASCA and BeppoSAX (Patel et al. 2004),
the September 2004 data correspond to an unabsorbed flux of
7.2$\times10^{-13}$ erg cm$^{-2}$ s$^{-1}$ (2-10 keV). This is a
factor five below  that seen with the above satellites in 1999,
and the smallest flux detected from \igr. For this spectrum, the
February 2004 observation, yields a 3$\sigma$ upper limit of
$\sim4\times10^{-14}$ erg cm$^{-2}$ s$^{-1}$  on the 2-10 keV
absorbed flux.  For an assumed distance of 7 kpc, the luminosity
upper limit is $\sim4\times10^{32}$ erg s$^{-1}$.

\begin{figure}[th!]
\begin{center}
\psfig{figure=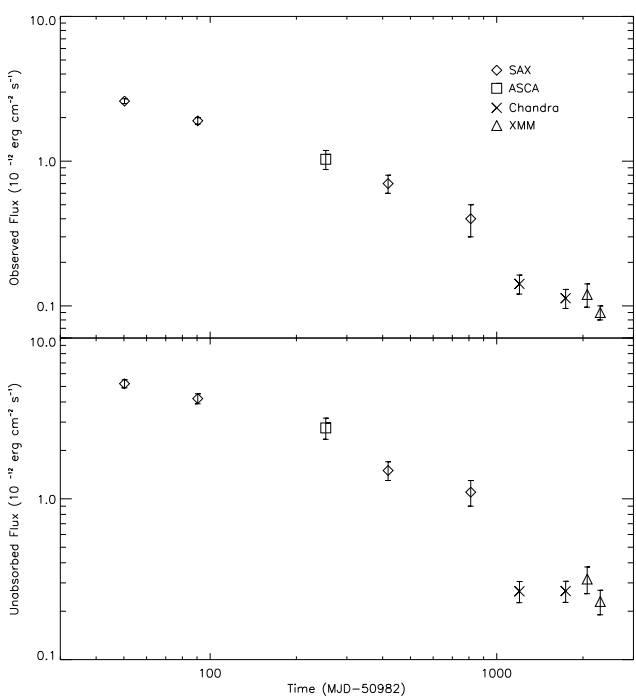,width=9.5cm,angle=90}

\caption{Long term light curve of \sgr\  based on data from
different satellites.  Top panel: absorbed flux in the 2-10 keV
range. Bottom panel: unabsorbed flux in the 2-10 keV range. The
ASCA and Chandra values have been computed from the spectral
parameters and unabsorbed fluxes given in Kouveliotou et al.
(2003). For clarity, the XMM-Newton points of September 4, which
are consistent with the last measurement, are not plotted. }
\end{center}
\end{figure}

\subsection{Reanalysis of the BeppoSAX observations of \sgr}

We have reanalyzed the four  observations of \sgr\ carried out
with the BeppoSAX satellite from August 1998 to September 2000. We
extracted the source spectra only from the Medium Energy
Concentrator Spectrometer (MECS) data\footnote{the Low Energy
Concentrator Spectrometer (LECS) data do not add significant
information, since \sgr\ was too absorbed and faint for this
instrument} using circular extraction regions with radius 2$'$.
This relatively small region was chosen in order  to reduce the
uncertainties related to the background subtraction, since the
source lies in a low galactic latitude field affected by the
presence of some diffuse emission and other confusing sources. The
background spectra were extracted from source free regions of the
same observations. The background subtracted spectra in the 1.8-10
keV energy range were rebinned in order to have at least 30 counts
per bin and fitted with an absorbed power law. The best fit
parameters are reported in Table 1. The small differences with the
results reported by Kouveliotou et al. (2003) are probably due to
different background estimation techniques.

\section{Discussion}

The light curve of \sgr\ based on data from the different
satellites is shown in Fig.~2, where the 2-10 keV flux values
correspond to the fits with the same absorption in all the
observations (N$_H$=9$\times10^{22}$ cm$^{-2}$). The long term
decrease in luminosity is clear, but, owing to the source spectral
variations,  the detailed shape of the decay is different for the
observed (upper panel) and unabsorbed (lower panel) flux.
Kouveliotou et al. (2003) fitted the decay of the unabsorbed flux
using a model involving a deep crustal heating following the 1998
bursting activity and requiring a massive neutron star (M$>$1.5
$M{_\odot}$). In particular, they pointed out that this model
could well explain the plateau between days 400 and 800, but
noticed that the March 2003 Chandra observation could not be
explained in this framework, suggesting that the source reached a
steady low level luminosity. According to our reanalysis of the
BeppoSAX data the evidence for a plateau between days 400 and 800
is not so compelling. In fact all the BeppoSAX and ASCA points,
before the rapid decline seen with Chandra in September 2001, are
well fit by a power law decay, F(t)$\propto$(t-t$_0$)$^{-\delta}$.
Fixing t$_0$ at the time of the discovery outburst, we obtain
$\delta$=0.6.

If one considers the observed fluxes, the Chandra and XMM-Newton
data suggest that \sgr\ has continued to fade also after September
2001. There is evidence that the spectrum softened between the two
Chandra observations (Kouveliotou et al. 2003). The photon index
measured with XMM-Newton is consistent with that of the last
Chandra observation but, due to the large uncertainties,  also a
further softening cannot be excluded. This apparent fading is not
necessarily related to a variation of the source overall
luminosity, as clearly indicated by the fluxes corrected for the
absorption plotted in the lower panel of Fig.~2.

The XMM-Newton data of September 2004 imply a luminosity
$\sim$3.5$\times$10$^{33}$d$^2_{11}$ erg s$^{-1}$. This is the
lowest luminosity observed from a SGR. The fact that \sgr\ has not
emitted bursts during the last $\sim$6 years suggests that a
luminosity below 10$^{34}$ erg s$^{-1}$ might be typical of
''quiescent'' SGRs. This simple interpretation is possibly
contradicted by the two following considerations. First, the SGR
in the Large Magellanic Cloud,  SGR 0526--66,  has a higher
luminosity (10$^{36}$ erg s$^{-1}$), but has not shown signs of
strong bursting activity in the last 15 years. However, faint
bursts, like those recently observed from SGR 1806--20 with
INTEGRAL (G\"{o}tz et al. 2004) might have passed undetected in
SGR 0526--66 due to its larger distance and location in a less
frequently monitored sky region. Second, most Anomalous X-ray
Pulsars (AXPs, see, e.g. Mereghetti et al. 2002 for a review),
which are also generally thought to be magnetars, have nearly
steady luminosity larger than 10$^{35}$ erg s$^{-1}$. Although
bursts have been observed in three of them (1E 1048.1--5937,
Gavriil et al. 2002; 1E 2259$+$586, Kaspi et al. 2003; XTE
J1810--197, Woods et al. 2005), there are a few AXPs which have
not shown any bursting activity and yet are relatively luminous
X-ray sources.

The low luminosity and soft spectrum of \sgr\ seen with XMM-Newton
are quite similar to the values measured in archival data of the
AXP XTE J1810--197 (Gotthelf et al. 2004) obtained before its
discovery as a bright transient source in January 2003 (Ibrahim et
al. 2004). Based on the currently available data, the two sources
seem to behave in a similar way. Further observations of XTE
J1810--197 will establish if and how a steady quiescent level is
attained.

We finally note that another soft repeater, SGR 1900+14, has
possibly been quiescent in the last three years. To our knowledge,
the last reported burst activity from this source occurred in
November 2002 (Hurley et al. 2002). Therefore it will be
interesting to see whether also in SGR 1900+14 the X-ray
luminosity will evolve toward a low state similar to that observed
for \sgr .

\begin{acknowledgements}

This work has been partially supported by the Italian Space Agency
and by the MIUR under grant PRIN 2004-023189.

\end{acknowledgements}

\end{document}